\documentclass[longbibliography,aps,prx,amsmath,amssymb,superscriptaddress,twocolumn,10pt]{revtex4-2}

\usepackage{geometry}
\usepackage{float}
\usepackage{algorithm}
\usepackage{algpseudocode}
\floatstyle{ruled}
\restylefloat{algorithm}
\usepackage[table,dvipsnames]{xcolor}
\usepackage{amsmath,amssymb}
\usepackage{graphicx}
\usepackage{natbib}
\usepackage{mathtools}
\usepackage{appendix}
\usepackage[colorlinks=true]{hyperref}
\usepackage{comment}
\usepackage{minted}
\usepackage{url}
\usepackage{blindtext}
\usepackage{siunitx}
\usepackage{microtype}
\usepackage{booktabs}
\usepackage[math]{cellspace}
\hypersetup{
  colorlinks = true,
  linkcolor  = blue,
  citecolor  = blue,
  urlcolor   = blue
}

\usepackage[utf8]{inputenc}
\usepackage[T1]{fontenc}
\usepackage{multirow}
\usepackage{tabularx}
\usepackage{bbm}
\usepackage{amsmath}
\usepackage[shortlabels]{enumitem}
\newcommand{\sign}[1]{\text{sgn}\left({#1}\right)}

\usepackage{newtxtext}
\usepackage{newtxmath}
\usepackage{float}

\usepackage{cleveref}
\crefname{equation}{Eq.}{Eqs.}
\crefname{section}{Sec.}{Sec.}
\crefname{figure}{Fig.}{Figs.}
\crefrangeformat{equation}{Eqs. (#3#1#4)-(#5#2#6)}
\crefrangeformat{section}{Sec. #3#1#4-#5#2#6}
\crefrangeformat{figure}{Figs. #3#1#4-#5#2#6}
\crefname{appendixsection}{Appendix}{Appendix}
\crefrangeformat{appendixsection}{Appendix #3#1#4-#5#2#6}

\usepackage[normalem]{ulem}

\begin{document}
\title{High-fidelity inference of power grid frequency distributions}
\date{today}

\author{Alessandro Lonardi}
\affiliation{Centre for Complex Systems, School of Mathematical Sciences, Queen Mary University of London, Mile End Road, London E1 4NS, United Kingdom}
\author{Benjamin Schäfer}
\affiliation{Karlsruhe Institute of Technology, Hermann-von-Helmholtz-Platz 1, 76344 Eggenstein-Leopoldshafen, Germany}
\affiliation{Helmholtz AI}
\author{Christian Beck}
\affiliation{Centre for Complex Systems, School of Mathematical Sciences, Queen Mary University of London, Mile End Road, London E1 4NS, United Kingdom}

\date{\today}

\begin{abstract}
Precision measurements of power grid frequency track energy supply-demand imbalances, with persistent fluctuations indicating strain that can lead to outages. Characterizing the statistical properties of grid frequency fluctuations is therefore essential to achieve efficient control and stable operations. Existing models had limited success in reconstructing the complex non-Gaussian features of observed frequency distributions. Here, we introduce a method for statistical inference of an interpretable stochastic process governing frequency fluctuations, modeling frequency through a coarse-grained power imbalance signal combined with nonlinear generator control and Gaussian white noise. We develop an efficient algorithm to infer these latent variables via maximum likelihood estimation, combined with superstatistics. We test our method on a large dataset of new measurements from Great Britain and South Africa. Although these grids have markedly different properties, our predictions for frequency distributions match measurements excellently. Our method uses only frequency data for inference, thus offering an alternative to data-intensive approaches.
\end{abstract}

\maketitle

\section{Introduction}
\label{sec:intro}

Global access to electricity rests on the reliable operation of power grids and, as such, is a problem in complexity science and dynamical systems \cite{anvari2026sustainability}. Indeed, power grids are large-scale dynamical networks in which synchronous operation must be continuously maintained to balance power generation and load. This balance requires that fluctuations from the nominal grid frequency remain small, typically of the order of one-tenth of a Hertz \cite{rebours2007survey}. When frequency bounds are violated, this often indicates that the power generation balance has been disrupted, for instance, due to a power shortage. In such situations, power grids experience stress that can escalate into outages with disastrous consequences, as recently seen in the Iberian Peninsula and Chile in 2025  \cite{iberianblackout,chileblackout}, and in Pakistan in 2023 \cite{pakistanblackout}.

At a coarse scale, the stochastic dynamics of grid frequency is straightforward: if there is an excess demand for electricity, rotors in synchronous machines convert mechanical power into electrical power and the frequency decreases; conversely, if there is a demand deficiency, electrical energy is converted to additional rotational energy and the frequency increases \cite{kundur2007power,machowski2020power}. At a finer scale, however, frequency behavior depends on many factors, such as renewable energy profiles, consumer demand, and energy market prices \cite{kruse2021revealing}. The interplay of these factors renders frequency dynamics epistemically uncertain, as it is difficult to determine their relations in large-scale systems. Despite this complication---and perhaps suprisingly---grid frequency exhibits non-Gaussian patterns leading to the emergence of multimodal, heavy-tailed distributions observed in experimental measurements across grids worldwide \cite{weissbach2009high,kashima2015modeling,schaefer2018non,schafer2018isolating,vorobev2019deadbands,anvari2020stochastic,gorjao2020open,gorjao2020data,delgiudice2021effects,wen2023non,kraljic2023towards,gorjao2023stochastic,gorjao2021spatio,onksaker2023predicting,kruse2023physics,oberhofer2023non,maritz2024data,wen2024identifying,drewnick2025analyzing,openacess2025grid,oberhofer2025nonlinear,lonardi2026understanding,wen2026power,hao2026stochastic}. Characterizing precisely such patterns gives insight into the bulk behavior of power grids and, of practical relevance to grid operators, enables accurate forecasting of their frequency
\cite{kruse2020predictability,onksaker2023predicting,kruse2023physics}.

Useful among approaches to capture the stochastic nature of grid frequency are statistical physics methods, which typically posit energy conservation via a stochastic differential equation governing frequency dynamics \cite{ulbig2014impact}. For instance, heavy tails in both frequency \cite{schaefer2018non} and frequency-increment \cite{gorjao2021spatio} distributions can be fitted via superstatistical models \cite{beck2001dynamical,beck2003superstatistics,beck2005from}, which assume that the frequency dynamics evolves through a succession of locally stationary states whose parameters fluctuate on long timescales. Superstatistics can also be extended to numerically simulate \cite{hao2026stochastic} and fit \cite{lonardi2026understanding} multimodal \emph{and} heavy-tailed frequency distributions.
While not explicitly superstatistical, time-dependent parameters controlling grid-frequency dynamics have been widely employed in numerical simulations \cite{weissbach2009high,vorobev2019deadbands,anvari2020stochastic,gorjao2020data,delgiudice2021effects,oberhofer2023non,kraljic2023towards}. The idea of this line of work is to combine a time-dependent function that describes the net power imbalance between generation and load with a stationary function that represents the control actions of grid generators to damp frequency fluctuations around the nominal value. Recently, this has led to simulations that correctly reproduce both frequency distributions and autocorrelation \cite{kraljic2023towards}.
An alternative perspective is offered by data-driven approaches, where physically motivated functional forms for the frequency-dynamics coefficients are made state-dependent to reproduce distributional features, though still relying on simulations \cite{gorjao2023stochastic,oberhofer2023non,oberhofer2025nonlinear}.

Despite substantial progress, a clear knowledge gap remains: no existing approach achieves high-accuracy statistical inference of frequency probability distributions. That is, a faithful reconstruction of measured data via principled estimation of their underlying distribution. Current inference approaches remain either coarse-grained \cite{schaefer2018non,schafer2018isolating,gorjao2021spatio,hao2026stochastic} or limited to restricted regions of the distributions \cite{lonardi2026understanding}. Simulations, by contrast, may accurately reproduce statistical features. Still, their agreement with data is typically assessed a posteriori rather than being enforced as an a priori criterion for parameter identification.

\begin{figure}[t]
    \centering
    \includegraphics[width=1\linewidth]{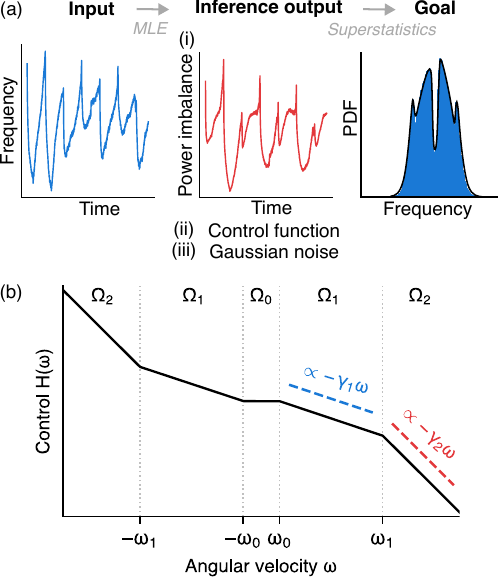}
    \caption{(a) Methodology overview. MLE is performed using measured frequency time series as the sole input. MLE jointly infers the power imbalance, control function, and Gaussian white noise intensity. The inferred quantities are then combined in a superstatistical model to fit the measured frequency distribution. (b) Piecewise-linear control function as in \Crefrange{eq: control 1}{eq: control 2}. Within the frequency deadband $\Omega_0$, control is inactive. Outside $\Omega_0$, in $\Omega_1$ and $\Omega_2$, frequency deviations are linearly damped by the coefficients $\gamma_2 \geq \gamma_1$, driving the frequency back toward its nominal value.}
    \label{fig: diagram}
\end{figure}

Here, we fill this gap by formulating a statistical inference problem with a physically interpretable low-dimensional latent structure directly on the frequency dynamics. Specifically, we assume that the frequency dynamics is governed by a slow, superstatistical power imbalance that drives the grid's long-term evolution, a nonlinear control function representing the damping effect of generators, and Gaussian white noise. We jointly infer these contributions using maximum likelihood estimation (MLE), which we solve by developing and implementing a block-coordinate descent algorithm \cite{beck2017first} at low computational cost. Once the latent structure of the frequency dynamics is inferred, we leverage insights from superstatistical modeling \cite{lonardi2026understanding} and use them to fit frequency distributions with unprecedented accuracy. Our methodology is schematically illustrated in \Cref{fig: diagram}(a).

Alternative approaches for estimating latent parameters of frequency dynamics have also been explored, but they tend to be more restrictive than ours. For instance, to estimate generators' damping coefficients in isolation, without accounting for the power imbalance, one can employ kernel regression on frequency time series \cite{gorjao2019kramersmoyal,gorjao2020open,gorjao2020data,gorjao2023stochastic,wen2023non,maritz2024data,drewnick2025analyzing,oberhofer2025nonlinear,lonardi2026understanding}, use machine learning methods such as gradient-boosted trees \cite{drewnick2025analyzing}, directly fit the frequency autocorrelation \cite{schaefer2018non,lonardi2026understanding}, or leverage first-passage processes \cite{lonardi2026understanding}.

For joint estimation, a flexible method is Sparse Identification of Nonlinear Dynamics (SINDy) \cite{wen2024identifying,wen2026power}, which identifies the dynamical equations that best fit the frequency dynamics from a predefined library. In settings where detailed data on the socio-technical parameters influencing power consumption are available, such as Europe \cite{entsoe2020}, latent parameters can also be learned via backpropagation in physics-informed neural networks \cite{kruse2023physics}.

Unlike the latter data-intensive neural networks, our method has the advantage of operating in data-scarce settings, sometimes corresponding to comparatively understudied power grids. In fact, we perform inference only on frequency time series, without requiring any additional data. We make a case in point by testing the algorithm on large-scale frequency time series measured in London, Great Britain (GB; we use GB instead of UK since Northern Ireland has its own grid), and Stellenbosch, South Africa (SA). These two grids are similar in that they are central to their respective energy markets \cite{epexukmarket,sappmarket}, but operate under markedly distinct conditions \cite{lonardi2026understanding} and yield substantially different frequency distributions. Yet, our approach achieves excellent accuracy in both cases.

Our algorithm \cite{codefolder} and data \cite{datafolder} are fully open-sourced.

\section{Continuous-time stochastic grid dynamics}
\label{sec: continuous-time stochastic grid dynamics}

\subsection{Aggregated swing equation}

The stochastic dynamics of grid frequency is our starting point. For continuous times, the bulk behavior of a grid can be described by the aggregated swing equation \cite{ulbig2014impact}. This equation expresses energy conservation in terms of the bulk angular velocity $\omega(t)  = 2 \pi (f(t) - f_\mathrm{R})$ (expressed in $\mathrm{rad/s}$), which measures deviations of the frequency $f(t)$ from its nominal value $f_{\mathrm{R}} = \qty{50}{Hz}$ (or $\qty{60}{Hz}$ in some parts of the world, e.g., the US, Canada, South Korea). We write it as
\begin{equation}
    \label{eq: aggregated swing equation}
    \frac{\mathrm{d} \omega}{\mathrm{d} t } = H(\omega; \gamma_1, \gamma_2) + P(t) + \epsilon \xi(t) \,.
\end{equation}

In \Cref{eq: aggregated swing equation}, $P(t)$ is the power imbalance, capturing mismatches between power generation and load arising, for instance, from aggregate consumer behavior or energy market transactions \cite{schaefer2018non,schafer2018isolating,gorjao2020data,boettcher2026impact,lonardi2026understanding}.

The control function $H(\omega; \gamma_1, \gamma_2)$ damps fluctuations and brings $\omega(t)$ back to zero after a disturbance. We model it as a piecewise linear function depending on two coefficients $\gamma_2 \geq \gamma_1$, as shown in \Cref{fig: diagram}(b). In formulas,
\begin{alignat}{2}
    \label{eq: control 1}
    H(\omega) &= - \sign{\omega} h (|\omega|) \\
    \label{eq: control 2}
    h(\omega) &=
        \begin{cases}
        0 \quad & \omega \in \Omega_0\\
        \gamma_1 (\omega - \omega_0) \quad & \omega \in \Omega_1 \\
        \gamma_2 (\omega - \omega_1) + \gamma_1 (\omega_1 - \omega_0) \quad & \omega \in \Omega_2 \,,
        \end{cases}
\end{alignat}
where $\Omega_0 = [0, \omega_0)$, $\Omega_1 = [\omega_0, \omega_1)$, and $\Omega_2 = [\omega_1, + \infty)$. Piecewise-linear control is compatible with the grid codes implemented in Great Britain \cite{nesonote,ofgemnote} and SA \cite{nersacode}. In Great Britain, control linearly damps deviations in the so-called ``dynamic moderation'' and ``dynamic regulation'' regions $\Omega_1$ and $\Omega_2$. Hence, \Crefrange{eq: control 1}{eq: control 2} faithfully reproduce nominal regulations. In SA, generator responses are heterogeneous: no linear damping is mandated in $\Omega_1$, and individual units activate at different thresholds. Still, \Crefrange{eq: control 1}{eq: control 2} successfully approximates the staggered control response of generators \cite{lonardi2026understanding}. The narrow region $\Omega_0$ is a frequency deadband arising since generators can activate up to finite precision, typically of the order of $\qty{10}{mHz}$ \cite{rebours2007survey}. Following national grid codes, we set $\omega_0 = 2 \pi \cdot \qty{0.015}{rad/s}$ and $\omega_1 = 2 \pi \cdot \qty{0.1}{rad/s}$ in the UK, and $\omega_1 = 2 \pi \cdot \qty{0.15}{rad/s}$ in SA. For $\omega_0$ in SA, we are not aware of any nominal specification. Since experimental data do not provide significant evidence of a deadband \cite{lonardi2026understanding}, we set $\omega_0 = \qty{0}{rad/s}$.

Finally, $\xi(t)$ denotes Gaussian white noise with amplitude $\epsilon > 0$ accounting for fast frequency fluctuations that cannot be resolved by power imbalance or control. More precisely, the noise is the distributional derivative of a standard Wiener process $W(t)$ whose infinitesimal increments satisfy $\mathrm{d} W(t) \sim \mathcal{N}(0,\mathrm{d}t)$. Following previous studies \cite{gorjao2020data,lonardi2026understanding}, we use Gaussian white noise for greater analytical tractability. We find that such a choice does not affect the model's ability to reproduce the frequency distribution at high granularity, provided it is combined with nonlinear control and a slow-varying power imbalance as per \Cref{eq: aggregated swing equation} (see \Cref{ssec: fit of the frequency histograms}). Other choices, such as Lévy \cite{schaefer2018non,kashima2015modeling,schafer2018isolating} and fractional Gaussian or fractional Lévy noise \cite{kraljic2023towards}, are also possible. However, they complicate the analytical treatment.

Putting all variables of \Cref{eq: aggregated swing equation} together, given a trajectory $\omega(t)$, we want to infer the latent time-dependent power imbalance $P(t)$ and the scalar parameters $\boldsymbol{\theta} = (\gamma_1, \gamma_2, \epsilon)$.

\subsection{Superstatistical modeling}
\label{sec: superstatistical modeling}

Assuming for now that inference can be performed, we require a method to combine the inferred parameters to fit the measured frequency histograms. To that end, we employ superstatistical modeling \cite{beck2001dynamical,beck2003superstatistics,beck2005from}, particularly building on recent advances \cite{lonardi2026understanding}. The idea is to treat the power imbalance $P(t)$ as slowly varying or constant on short time intervals (think of seconds), since its typical variations occur on longer timescales (think of minutes) than the characteristic damping time of the control dynamics \cite{anvari2016short,kruse2023physics,lonardi2026understanding} (see \Cref{ssec: model validation} for a validation on GB and SA measurements).

Under this assumption, for short intervals, \Cref{eq: aggregated swing equation} reduces to an Ornstein--Uhlenbeck process with fixed drift and noise, yielding a Fokker--Planck equation for a conditional distribution $f(\omega \,|\, P)$ of frequency deviations ``$\omega$-given-$P$''. The full stationary distribution of $\omega$ is then obtained as a superposition over all possible values of $P$,
\begin{equation}
    \label{eq: frequency distribution}
    p(\omega) = \int f(\omega \,|\, P) \, \varphi(P) \, \mathrm{d}P \,,
\end{equation}
where $\varphi(P)$ denotes the power imbalance distribution.

Assuming Gaussian white noise, $f(\omega \,|\, P)$ admits a closed-form expression up to normalization \cite{lonardi2026understanding} (see \Cref{sec: closed form expression of f}). Thus, once $P(t)$ is inferred, its distribution $\varphi(P)$ can be estimated and, together with $\boldsymbol{\theta}$, used to numerically evaluate the integral in \Cref{eq: frequency distribution}, thereby fitting measurements.

This procedure leads to the missing modeling step: building a method to infer $P(t)$ and $\boldsymbol{\theta}$ from frequency data.

\section{Inference}
\label{sec: inference}

\subsection{Model discretization}
\label{ssec: model discretization}

We measure high-resolution frequency signals using phasor measurement units (PMUs)
\footnote{Technical details on the PMUs and on the data acquisition process are detailed in \citet{lonardi2026understanding}}.
Each PMU returns a series $\boldsymbol{\omega} = (\omega_0, \dots, \omega_k, \dots, \omega_{T-1}) \in \mathbb{R}^T$
whose entries are measurements at time $t_k$, i.e., $\omega_k = \omega(t_k)$. Data are recorded at fixed intervals $\Delta t = t_{k+1}-t_k$.

On the PMUs temporal grid, we discretize \Cref{eq: aggregated swing equation} with the Euler-Maruyama method \cite{kloeden1977numerical}:
\begin{equation}
    \label{eq: euler maruyama discretization}
    \boldsymbol{\Delta  \omega} = \Delta t \, (\mathbf{H}+ \mathbf{P} ) + \sqrt{\Delta t} \, \epsilon \boldsymbol{\xi} \,.
\end{equation}
Here, $\boldsymbol{\Delta \omega} \in \mathbb{R}^{T-1}$ 
collects all finite differences $\Delta \omega_k = \omega_{k+1} - \omega_k$ and, accordingly, $\mathbf{P} \in \mathbb{R}^{T-1} $, $\mathbf{H} \in \mathbb{R}^{T-1} $, and $\boldsymbol{\xi} \in \mathbb{R}^{T-1} $ express the power imbalance, the control function, and the noise statistics at each time step. In $\mathbf{H}$, we omit writing $\gamma_1$ and $\gamma_2$ for brevity and $\xi_k \sim \mathcal{N}(0,1)$ are i.i.d.

While formally consistent, \Cref{eq: euler maruyama discretization} allows the power imbalance $\mathbf{P}$ to vary at the same temporal resolution as $\boldsymbol{\Delta \omega}$. This is undesirable, as it leads to an inference algorithm in which all fluctuations in the data are explained by the power imbalance, leaving control and noise idle. We address this issue by constraining the power imbalance to a coarse grid with step size $N \Delta t$, where $N$ is a downsampling factor. To obtain physically plausible and accurate results, we select $N$ via cross-validation, identifying a regime in which $N$ is large enough that the inferred noise amplitude $\epsilon$ does not collapse to zero due to the power imbalance absorbing fast fluctuations, but not so large that the coarse-graining obscures slow trends in the frequency data that drive the control response (see \Cref{ssec: inference results} for experimental results on GB and SA measurements).

The resulting coarse grid has $M$ points. Therefore, the coarse power imbalance is a low-dimensional vector $\tilde{\mathbf{P}} \in \mathbb{R}^M$ whose entries $\tilde{P}_j$ vary every $N$ frequency measurements. To reconstruct a fine-grid power signal compatible with the frequency dynamics, we interpolate back the coarse-grid values onto the fine grid with a linear interpolation matrix $\mathbf{B} \in \mathbb{R}^{(T-1) \times M}$ such that
\begin{equation}
    \label{eq: interpolation}
    \mathbf{P} = \mathbf{B}\tilde{\mathbf{P}} \,.
\end{equation}
Each row of $\mathbf{B}$ corresponds to a fine-grid time index $k$ and expresses $P_k$ as a convex combination of its two neighboring coarse-grid values $\tilde{P}_{j(k)}$ and $\tilde{P}_{j(k)+1}$. By construction, $\mathbf{B}$ is sparse, with at most two nonzero entries per row.

Substituting \Cref{eq: interpolation} into \Cref{eq: euler maruyama discretization} gives
\begin{equation}
    \label{eq: vectorized model coarse}
    \boldsymbol{\Delta  \omega} = \Delta t  \, (\mathbf{H}+ \mathbf{B} \tilde{\mathbf{P}} ) + \sqrt{\Delta t} \, \epsilon \boldsymbol{\xi} \,.
\end{equation}

\subsection{Likelihood}

The discretization in \Cref{eq: vectorized model coarse} readily yields the data likelihood. In fact, we can write the conditional distribution of measurement increments given the latent parameters as
\begin{equation}
    \label{eq: likelihood}
    p(\boldsymbol{\Delta \omega} \mid \tilde{\mathbf{P}}, \boldsymbol{\theta}) = \mathcal{N}(\Delta t \, (\mathbf{H} + \mathbf{B} \tilde{\mathbf{P}}), \epsilon^2 \Delta t \, \mathbf{I}) \,,
\end{equation}
where $\mathbf{I} \in \mathbb{R}^{{(T-1)} \times {(T-1)}}$ is the identity matrix. 

\Cref{eq: likelihood} gives the optimization objective we use for inference, as, up to additive constants independent of model parameters, it yields the negative log-likelihood
\begin{equation}
    \label{eq: mle objective}
    \mathcal{L}(\tilde{\mathbf{P}},\boldsymbol{\theta})
    =
    \frac{1}{2\epsilon^2\Delta t}
    \|
    \boldsymbol{ \Delta \omega}
    -
    \Delta t\, ( \mathbf{H} +\mathbf{B}\tilde{\mathbf{P}} )
    \|_2^2
    +
    \frac{T-1}{2}\log(\epsilon^2) \,,
\end{equation}
where $\| \cdot \|_2$ denotes the 2-norm. MLE then reduces to an optimization problem on \Cref{eq: mle objective}:
\begin{equation}
    \label{eq: optimization problem}
    \min_{\tilde{\mathbf{P}} \in \mathbb{R}^{M},\boldsymbol{\theta} \in \Theta}
    \mathcal{L}(\tilde{\mathbf{P}},\boldsymbol{\theta}) \,
\end{equation}
with $\Theta = \{(\gamma_1,\gamma_2,\epsilon)\in\mathbb{R}^3 \mid \epsilon>0,\;\gamma_2\geq\gamma_1>0\}.$

\subsection{block-coordinate descent}

To tackle \Cref{eq: optimization problem}, we develop a block-coordinate descent algorithm \cite{beck2017first}. Namely, we alternate between updating $\tilde{\mathbf{P}}$ while keeping $\boldsymbol{\theta}$ fixed and updating $\boldsymbol{\theta}$ with $\tilde{\mathbf{P}}$ fixed (see \Cref{sec: implementation details} for implementation details). As we show, our optimization algorithm is amenable to precomputation. In practice, it consistently achieves runtimes on the order of seconds across all our experiments.

\subsubsection{Least-squares for the power imbalance}
\label{sec: least-squares for the power imbalance}

We first discuss the update of $\tilde{\mathbf{P}}$ with $\boldsymbol{\theta}$ fixed. Here, minimizing \Cref{eq: mle objective} amounts to solving a least-squares problem since all terms depending only on $\boldsymbol{\theta}$ can be neglected. In particular, by defining the residual vector
\begin{equation}
    \label{eq: residual}
    \mathbf{r} = \boldsymbol{\Delta \omega} - \Delta t\,\mathbf{H}
\end{equation}
and the design matrix
\begin{equation}
    \label{eq: design matrix}
    \mathbf{A} = \Delta t\,\mathbf{B} \,,
\end{equation}
the least-squares objective yielded by \Cref{eq: mle objective} becomes
\begin{equation}
    \label{eq: least squares obj}
    \mathcal{J}(\tilde{\mathbf{P}} \mid \boldsymbol{\theta})
    =
    \|\mathbf{r}-\mathbf{A}\tilde{\mathbf{P}}\|_2^2 \,.
\end{equation}
MLE for $\tilde{\mathbf{P}}$ with $\boldsymbol{\theta}$ fixed can therefore be performed by solving the normal equations given by $\boldsymbol{\nabla} \mathcal{J} = \mathbf{0}$,
\begin{equation}
    \label{eq: normal equation}
    \mathbf{A}^T\mathbf{A}\,\tilde{\mathbf{P}}
    =
    \mathbf{A}^T\mathbf{r} \,,
\end{equation}
where the matrix $\mathbf{A}^T\mathbf{A}\in\mathbb{R}^{M\times M}$ is sparse (tridiagonal) and symmetric positive definite, hence invertible. This matrix depends only on $\mathbf{B}$ and can be precomputed prior to optimization. As a result, the updates of $\tilde{\mathbf{P}}$ amount to a sparse matrix-vector multiplication on the right-hand side of \Cref{eq: normal equation}, followed by the solution of a sparse linear system, which we perform efficiently with a sparse direct solver.

\subsubsection{MLE for the grid parameters}

To update $\boldsymbol{\theta}$ with $\tilde{\mathbf P}$ fixed, rather than optimizing naively over $\boldsymbol{\theta}$, we notice that \Cref{eq: mle objective} is convex in $\epsilon$ and admits a closed-form optimum for fixed $(\gamma_1,\gamma_2)$. We thus analytically find this optimum and substitute it into the negative log-likelihood, obtaining a profile likelihood that depends only on $(\gamma_1,\gamma_2)$. The parameter $\boldsymbol{\theta}$ is found by working with this reduced objective.

In formulas, defining the residual vector
\begin{equation}
\mathbf{e} 
=
\boldsymbol{\Delta\omega}
-
\Delta t\, ( \mathbf{H} + \mathbf{B} \tilde{\mathbf P} )
\end{equation}
and setting $\partial \mathcal{L} / \partial (\epsilon^2) = 0$ yields
\begin{equation}
    \label{eq: epsilon mle}
    \epsilon^{2}
    =
    \frac{\|
    \mathbf {e}
    \|_2^2}{(T-1)\Delta t}
     \,.
\end{equation}
Substituting \Cref{eq: epsilon mle} into \Cref{eq: mle objective} gives, up to negligible additive constants, the profile negative log-likelihood
\begin{equation}
\label{eq: profile likelihood}
{\mathcal{L}}_{\text{profile}}(\gamma_1,\gamma_2 \mid \tilde{\mathbf P})
=
\frac{T-1}{2}
\log
(
\|
\mathbf{e}
\|_2^2
) \,
\end{equation}
that we minimize numerically \cite{byrd1995limited}.
To enforce the constraints $\gamma_2 \geq \gamma_1 > 0$, we use the
log-parameterization $\alpha = \log \gamma_1$, $\beta = \log(\gamma_2-\gamma_1)$, and optimize over 
$(\alpha,\beta)$. At termination, we first recover $(\gamma_1,\gamma_2)$, then can compute $\epsilon$ once via \Cref{eq: epsilon mle}.

\section{Results}
\label{sec: results}

\subsection{Dataset}
\label{sec: dataset}

We consider frequency time series measured in London, UK, and Stellenbosch, SA. They span 8 months in 2025, from the 28th April to the 31st December, and are spaced by $\Delta t= \qty{1}{s}$, totaling approximately 43 million data points.

To run our algorithm, we split each time series into 496 batches of 12 hours, yielding a batch size of $T = 43200$ for each run. A small fraction of the data ($<0.05\%$) is unavailable due to gaps in PMU measurements. These missing values occur over 3 and 18 batches in GB and SA, respectively, for which we do not run the algorithm.

\subsection{Inference results}
\label{ssec: inference results}

\begin{figure*}[t]
    \centering
    \includegraphics[width=\linewidth]{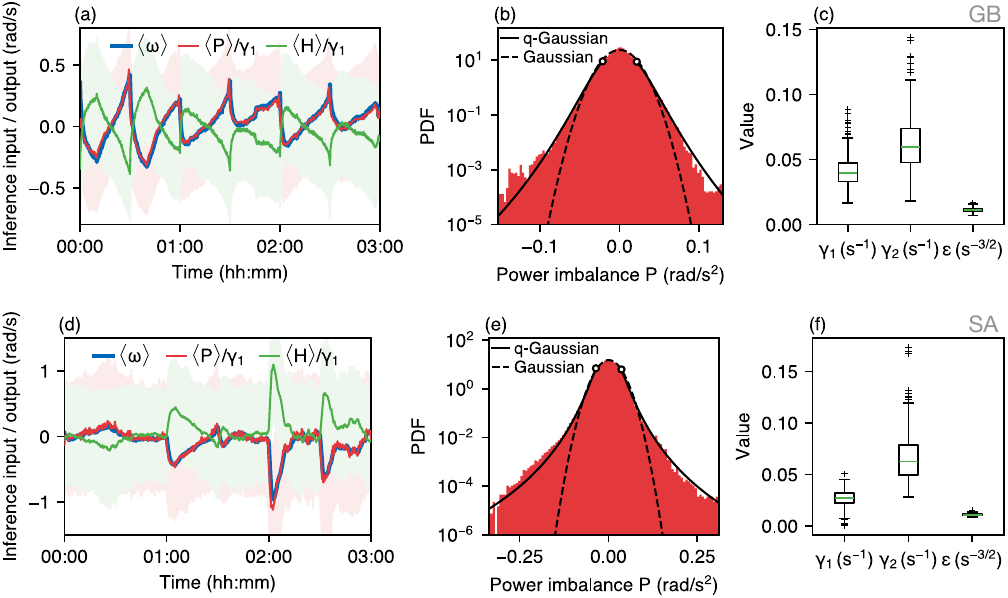}
    \caption{
    Inference results. Panels (a)-(c) are for GB and (d)-(f) are for SA. (a), (d) Three-hour snapshot of the daily-averaged frequency, power imbalance, and control. Shaded areas are standard deviations for power imbalance and control. We apply the rescaling $\langle P\rangle/\gamma$ and $\langle H\rangle/\gamma$, following the discussion in \Cref{ssec: inference results}. We set $\gamma=\gamma_1$ for convenience, selecting the values among those of (c) and (f) that best fit the frequency distributions in \Cref{fig: frequency fits} (see \Cref{ssec: fit of the frequency histograms} for details).
    (b), (e) Inferred power imbalance distribution. The plot is in semilogarithmic scale to highlight heavy tails. Gaussian fits are shown as dashed black lines and $q$-Gaussian fits for the tails as solid black lines.
    (c), (f) Boxplots of the inferred parameters $\boldsymbol{\theta}=(\gamma_1,\gamma_2,\epsilon)$. Green lines indicate medians, while boxes span the interquartile range.}
    \label{fig: inference results}
\end{figure*}

We show the inference results in \Cref{fig: inference results}. Here, we use downsampling parameters $N = 40$ and $N = 20$ for GB and SA, respectively. These values are selected via cross-validation to mitigate overfitting: $N$ should be large enough to prevent the inferred power imbalance from fitting measurement noise, yet sufficiently small to capture the slower variations present in the frequency time series (see \Cref{sec: mitigating overfitting} for details and numerical experiments). A smaller value of $N$ for SA than in GB is consistent with the more rapid temporal fluctuations observed in South African frequency measurements \cite{lonardi2026understanding}.

In \Cref{fig: inference results}(a), (d), we show a snapshot of the inferred power imbalance and control function (appropriately rescaled), together with the measured frequency profile, all averaged over the course of a day. The inferred power imbalance qualitatively follows the frequency. This behavior can be understood by approximating \Cref{eq: aggregated swing equation}. During regular operation, power grids spend most of their time near stable operating points where power imbalance and control contributions approximately cancel each other out. Averaging over daily batches also suppresses stochastic noise fluctuations, hence $\mathrm{d}\langle\omega\rangle/\mathrm{d}t = \langle H(\omega) \rangle + \langle P\rangle \approx 0$, $\langle\epsilon \xi\rangle = 0$, with $\langle \cdot \rangle$ denoting daily averaging. For linear control, $H(\omega) \approx -\gamma\omega$, this implies $\langle P\rangle \approx \gamma \langle \omega\rangle$, which is precisely the qualitative behavior observed in the data. In truth, the measurement increments $\boldsymbol{\Delta \omega}$ are structured, not simply roughly zero \cite{kraljic2023towards}, and their structure is precisely the information leveraged by our algorithm for inference. Still, this simple argument helps hone intuition and connects our results to numerical simulations in which the power imbalance is set equal to the averaged frequency profile as a proxy \cite{kraljic2023towards}.

In \Cref{fig: inference results}(b), (e), we plot the power imbalance distributions $\varphi(P)$. A striking feature is their heavy-tailedness.
Heavy tails are a distinctive signature of grid-frequency statistics worldwide \cite{gorjao2020data,schaefer2018non,schafer2018isolating,gorjao2021spatio,kashima2015modeling,weissbach2009high,kruse2023physics,anvari2020stochastic}. Under our model, heavy tails observed in the frequency distribution are inherited from the power imbalance itself and
disentangled from control, which instead yields Gaussian
tails in $f(\omega \mid P)$ (see \Cref{sec: closed form expression of f}). This interpretation was discussed theoretically \cite{lonardi2026understanding}, but so far lacked the experimental validation we provide (see \Cref{sec: tails of the frequency distributions} for an additional figure). We fit the tails, defined as the outermost 20\% of observations by absolute magnitude ($|P|$ above the 80th percentile), with a $q$-Gaussian, which yields power-law decays in the tails. By doing so, we find $q = 1.160 \pm 0.003$ in GB and $q = 1.273 \pm 0.002$ in SA. The non-extensivity parameter $0 < q < 3$ controls the heavy-tailedness of a $q$-Gaussian, with heavy tails appearing when $q > 1$. Both distributions are well fitted by values of $q$ in this range, with SA exhibiting the larger non-extensivity and correspondingly heavier tails (see \Cref{sec: tails of the frequency distributions} for code details).

\begin{figure}[b]
    \centering
    \includegraphics[width=0.85\linewidth]{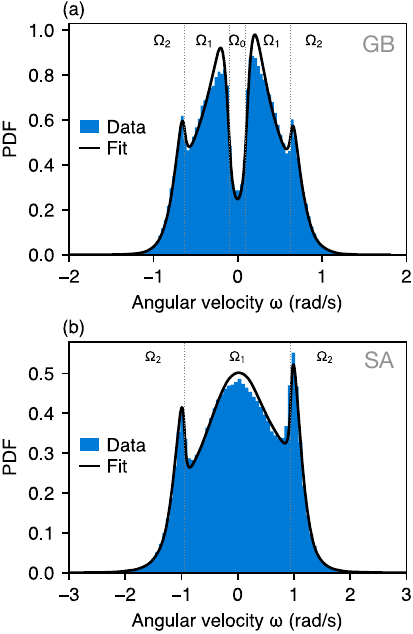}
    \caption{Frequency fits. Panel (a) is for GB and panel (b) for SA. Black lines are model predictions obtained via \Cref{eq: frequency distribution} while data histograms are in blue. Control regions mandated by countries' grid codes are separated by gray dotted vertical lines.}
    \label{fig: frequency fits}
\end{figure}

We also observe a slight asymmetry in the inferred power imbalance distribution in SA, with $\mathrm{skewness} \approx -0.16$. This asymmetry is reflected in the frequency distribution fit as a mismatch in the mode heights faithfully reproducing measured data, as shown in \Cref{fig: frequency fits}(b). In the UK, asymmetry is less evident. Still, the nontrivial structure of the power imbalance distribution is similarly imprinted in \Cref{fig: frequency fits}(a).

In our algorithm runs, we also slightly enlarged the GB frequency deadband to further reduce overfitting while introducing a small bias in the model (see \Cref{sec: biasing the inferred power signal} for a precise characterization).

Finally, in \Cref{fig: inference results}(c), (f), we show boxplots for the values of $\boldsymbol{\theta} = (\gamma_1, \gamma_2, \epsilon)$ inferred over the daily batches. We find that $\gamma_2$ typically attains larger values than $\gamma_1$, consistent with expectations for nominal control policies in both countries \cite{nesonote,ofgemnote,nersacode}. Larger damping values in ``outer'' control regions are key to producing multimodal frequency distributions \cite{lonardi2026understanding}, as we demonstrate experimentally in \Cref{fig: frequency fits}.

\subsection{Superstatistics validation}
\label{ssec: model validation}

Before fitting the frequency distributions, we verify that the superstatistical timescale separation holds for the inference outputs. Indeed, the integral in \Cref{eq: frequency distribution} is valid and can be rightfully computed only under the assumption that the power imbalance evolves on timescales longer than those of the fast control. 
This assumption is the foundation of the dynamical formulation of superstatistics \cite{beck2001dynamical,beck2005from} and, in practice, its validity is typically assessed by examining how data statistics such as skewness and kurtosis evolve over longer time windows \cite{beck2005from,vander2009superstatistical}, by extracting characteristic damping coefficients from the data autocorrelation \cite{lonardi2026understanding}, or by combining both approaches \cite{schaefer2018non}.

\begin{figure}[t]
    \centering
    \includegraphics[width=\linewidth]{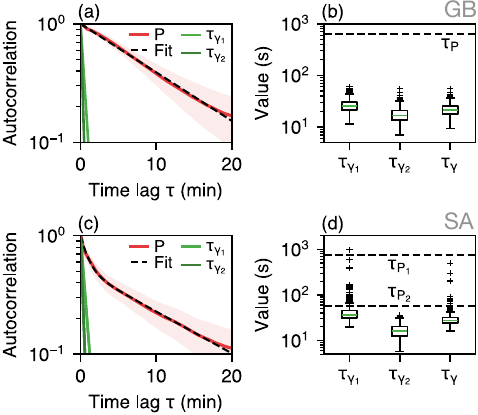}
    \caption{Superstatistics validation. Panels (a), (b) are for the UK, and panels (c), (d) are for SA. (a), (c) Autocorrelation decay of the power imbalance. Average values are shown in red, and their standard deviations are the shaded regions. These are computed over four-day-long data batches to reduce finite-size effects in the autocorrelation. Black dashed lines correspond to the exponential fits. We also plot two exponential decays $\exp(-\gamma_1 \tau)$ and $\exp(-\gamma_2 \tau)$, where the damping coefficients are those that best fit the frequency distributions in \Cref{fig: frequency fits} (see \Cref{ssec: fit of the frequency histograms} for details). (b), (d) Boxplots of $\tau_{\gamma_1}$, $\tau_{\gamma_2}$, and $\tau_{\gamma}$. Green lines indicate the median, while boxes span the interquartile range. Dashed lines indicate $\tau_P$ from panel (a) or $\tau_{P_1}$ and $\tau_{P_2}$ from panel (c).}
    \label{fig: validation}
\end{figure}

Here, we compute the autocorrelation of the inferred power imbalance and extract characteristic timescales over which it varies via an exponential fit. Control timescales are readily found with the inferred coefficients $\gamma_1$ and $\gamma_2$ as $\tau_{\gamma_1}=1/\gamma_1$, $\tau_{\gamma_2}=1/\gamma_2$, and the average $\tau_\gamma = (\tau_{\gamma_1}+\tau_{\gamma_2})/2$, evaluated across all batches. Results are shown in \Cref{fig: validation}.

In the UK, the autocorrelation of the power imbalance exhibits a clear exponential decay and thus is well-fitted by the ansatz $\mathrm{AC}(\tau) = \exp(-\tau / \tau_{P})$, where $\tau$ denotes the time lag, as shown in \Cref{fig: validation}(a). This empirical evidence also suggests that the inferred power imbalance can be modeled by a linear Markovian relaxation process with Gaussian white noise, consistent with previous numerical simulations \cite{vorobev2019deadbands}. More importantly, as it is evident in \Cref{fig: validation}(b), the fitted power imbalance timescale is well separated from the control timescales. In particular, 
\begin{align}
\tau_P &= \qty{637.5 \pm 0.6}{s} \\
\tau_{\gamma_1} &\approx \qty{25.4}{s}  \quad \tau_{\gamma_2} \approx  \qty{16.8}{s} \quad  \tau_{\gamma} \approx  \qty{21.2}{s} \,,
\end{align}
which roughly gives $\tau_P \approx 20 \cdot \tau_{\gamma}$ to support superstatistics. The reported control timescales are median values across batches, rounded to the same significant digits as $\tau_P$.

For SA, \Cref{fig: validation}(c) does not display a single exponential decay. Instead, the autocorrelation exhibits two timescales $\tau_{P_1}$ and $\tau_{P_2}$ that we find by fitting $\mathrm{AC}(\tau) = A \exp(-\tau / \tau_{P_1}) + (1-A)\exp(-\tau / \tau_{P_2})$, where $0 \leq A \leq 1$ is a scalar weight. Here, control and power imbalance timescales are generally more mildly separated, as shown in \Cref{fig: validation}(d), and give
\begin{align}
\tau_{P_1} &= \qty{58.5 \pm 0.4}{s} \quad \tau_{P_2} = \qty{740 \pm 2}{s} \\
\tau_{\gamma_1} &\approx \qty{37}{s}  \quad \tau_{\gamma_2} \approx \qty{16}{s} \quad  \tau_{\gamma} \approx \qty{27}{s} \,.
\end{align}
Therefore, the timescale separation is less pronounced when $\tau_{P_1}$ ($\tau_{P_1} \approx 2.2 \cdot \tau_{\gamma}$) is taken as reference. Still, it becomes clear relative to $\tau_{P_2}$ ($\tau_{P_2} \approx 27 \cdot \tau_{\gamma}$) and the hierarchy between power imbalance and control timescales is always preserved.

The double exponential decay observed in SA is consistent with previous empirical findings \cite{lonardi2026understanding}, but we interpret it differently here. In the literature, it is attributed to the distinct damping coefficients $\gamma_1$ and $\gamma_2$ of piecewise control. This interpretation is supported by long timescales of the order of hours and faster control timescales of the order of tens of minutes. 
Such timescales, while exhibiting a clear superstatistical separation, are, however, typically estimated independently of one another. In particular, long timescales are either assumed \cite{lonardi2026understanding} or inferred from distributional statistics \cite{schaefer2018non}, whereas fast timescales are extracted from the frequency autocorrelation \cite{lonardi2026understanding,schaefer2018non}.

By contrast, our model jointly infers the power imbalance and the control damping coefficients, and in turn their timescales. By doing so, it separates slow and fast data fluctuations by design, as the inferred control accounts only for fluctuations that cannot be explained by the power imbalance itself. Previous studies have used proxies such as Gaussian filtering to remove slow fluctuations in the data before estimating control parameters \cite{gorjao2020data,drewnick2025analyzing,lonardi2026understanding}. While effective, such approaches remain heuristic compared to a likelihood-based method. As a result, we find that the double decay observed in SA originates from the inferred power imbalance itself, while the characteristic control timescales are of the order of seconds. We note that this argument does not invalidate previous literature, as control parameters should always be understood as effective quantities that depend on the timescale separation.

For our model in particular, we also find that the superstatistical timescale validation is dependent on the downsampling factor $N$. We find that the long decay timescale of the power imbalance is largely insensitive to $N$, but $N$ evidently influences the inferred control coefficients indirectly through the inferred power imbalance and, in turn, modifies the characteristic control timescales (see \Cref{sec: mitigating overfitting}).

\subsection{Fit of the frequency distributions}
\label{ssec: fit of the frequency histograms}

We finally fit the frequency distributions by integrating \Cref{eq: frequency distribution} and compare model predictions with measured data in \Cref{fig: frequency fits}. Our results are remarkably accurate and achieve higher fidelity compared to previous maximum-likelihood approaches \cite{schaefer2018non,schafer2018isolating,gorjao2021spatio,hao2026stochastic} and numerical simulations \cite{vorobev2019deadbands,anvari2020stochastic,gorjao2020data,delgiudice2021effects,oberhofer2023non,kraljic2023towards} (see \Cref{sec: tails of the frequency distributions} for a quantitative comparison with standard Gaussian and $q$-Gaussian fits applied to GB and SA measurements).

In particular, both fits precisely capture the nontrivial multimodal structure of the distributions, including peaks of different magnitude and location, asymmetries inherited from $\varphi(P)$ (as discussed in \Cref{ssec: inference results}) as well as heavy tails (see \Cref{sec: tails of the frequency distributions} for a figure).

In GB, as shown in \Cref{fig: frequency fits}(a), the model reproduces the deadband dip in $\Omega_0$ caused by idle control \cite{lonardi2026understanding}, as well as secondary peaks at the transition between ``dynamic moderation'' and ``dynamic regulation'' in $\Omega_1$ and $\Omega_2$, where changes in nominal control are mandated \cite{nesonote,ofgemnote}. In SA, generators providing ``instantaneous reserve'' \cite{nersacode} activate in a staggered manner when frequency deviations remain moderate, i.e., within $\Omega_1$. At the boundary between $\Omega_1$ and $\Omega_2$, multiple generators activate simultaneously, producing sharp peaks that are accurately captured by our model.

A key factor shaping these peaks is the magnitude difference between the control coefficients $\gamma_1$ and $\gamma_2$ \cite{lonardi2026understanding}. In our experiments, inference is performed separately on daily batches, so multiple values of $\boldsymbol{\theta}$ are available (one per batch). To select optimal parameters to fit the data, we perform random-restart hill-climbing optimization over $\boldsymbol{\theta} = (\gamma_1, \gamma_2, \epsilon)$ \cite{russell1995modern}. Starting from a random initial batch, the algorithm sweeps through entries of $\boldsymbol{\theta}$ across batches and accepts those that improve the agreement between the model and the data, as quantified by the negative log-likelihood of the observed frequencies under the theoretical distribution. To avoid getting trapped in local minima, the procedure is periodically restarted when no improvement is observed for a fixed number of steps (see \Cref{sec: tails of the frequency distributions} for details).

\section{Discussion}
\label{sec: discussion}

In this work, we advanced the methodological foundations of statistical inference for power-grid frequency dynamics and validated our findings using new large-scale measurements collected in GB and SA.

Specifically, we developed a computationally efficient algorithm that performs statistical inference over physically interpretable latent variables governing grid frequency dynamics, namely the imbalance between power generation and load and the control response of grid generators (\Cref{sec: inference}). By combining statistical inference with recent developments in superstatistical modeling \cite{lonardi2026understanding}, we obtained high-fidelity predictions of measured frequency distributions (\Cref{fig: frequency fits}), achieving a level of agreement that surpasses previous inference-based approaches and numerical simulations alike \cite{schaefer2018non,schafer2018isolating,gorjao2021spatio,hao2026stochastic,vorobev2019deadbands,anvari2020stochastic,gorjao2020data,delgiudice2021effects,oberhofer2023non,kraljic2023towards}.

The accuracy of these results is noteworthy given the markedly different operating conditions, control regulations \cite{nesonote,ofgemnote,nersacode}, and market structures \cite{epexukmarket,sappmarket} of the British and South African power grids. This underscores the flexibility of our approach, which is a parsimonious model built from a small number of physically interpretable parameters rather than a data-intensive inference scheme.

Looking ahead, a natural next step is to go beyond fitting frequency distributions and investigate frequency temporal signatures. For instance, the autocorrelation of frequency time series is widely employed to gain insight into the bulk behavior of power grids \cite{schaefer2018non,gorjao2020data,anvari2020stochastic,kraljic2023towards,lonardi2026understanding,hao2026stochastic}. At long temporal lags, typically above 30 minutes, the autocorrelation exhibits a non-trivial structure, with recurrent peaks associated with energy market transactions, as well as a slower decay reflecting long-memory effects in the system. These features have been reproduced in numerical simulations that incorporate fractional noise in \Cref{eq: euler maruyama discretization} \cite{kraljic2023towards}, but it remains unclear whether this is the best-suited modeling assumption, given that fractional noise is not readily motivated by physical or electrical engineering insights.

A longer-term expansion of our research could be a Bayesian reformulation of our inference scheme, in which physically motivated priors for the power imbalance and control parameters would enter \Cref{eq: optimization problem}. This type of advanced modeling could enable more robust inference that naturally alleviates overfitting via the priors, thereby reducing the need for the methodological steps discussed in \Crefrange{sec: mitigating overfitting}{sec: biasing the inferred power signal}.

\section*{Acknowledgments}
The authors acknowledge funding by the UKRI-STFC grant UKRI467 {\emph{``Stability of the South African power grid---a statistical physics-based approach''}} and the Helmholtz Association's Initiative and Networking Fund through Helmholtz AI. This research utilised Queen Mary's Apocrita HPC facility, supported by QMUL Research-IT \href{http://doi.org/10.5281/zenodo.438045}{http://doi.org/10.5281/zenodo.438045.} 

\appendix
\crefalias{section}{appendixsection}

\section{Details on superstatistical modeling}
\label{sec: closed form expression of f}

Using superstatistical modeling, we assume that the power imbalance $P(t)$ can be regarded as constant over short time intervals where $P(t) = P$. In this way, \Cref{eq: aggregated swing equation} yields the Fokker--Planck equation
\begin{equation}
    \label{eq: fokker planck equation}
    \frac{\partial f}{\partial t} = - \frac{\partial}{\partial \omega} [ (H( \omega; \gamma_1, \gamma_2) + P) f ] + \frac{\epsilon^2}{2}\frac{\partial^2 f}{\partial \omega^2} \,
\end{equation}
for the conditional distribution $f(\omega, t \mid P)$. Imposing the stationarity condition in \Cref{eq: fokker planck equation}, namely $\partial f / \partial t= 0$, which--in the spirit of superstatistics--should be valid only over short time intervals, we recover a local equation with derivatives only in the state variable $\omega$. Integrating this equation twice yields
\begin{align}
    \label{eq: closed from f}
    f(\omega \mid P) &= \frac{1}{Z} \exp( - \Phi(\omega \mid P)) \\
    \Phi(\omega \mid P) &= -\frac{2}{\epsilon^2} \int (H(\omega ; \gamma_1,\gamma_2) + P) \, \mathrm{d} \omega \,.
\end{align}

\Cref{eq: closed from f} cannot be written in closed form since the partition function $Z(P)$ cannot be integrated analytically. However, the potential $\Phi(\omega \mid P)$ can be calculated analytically and one obtains
\cite{lonardi2026understanding}
\begin{align}
    \label{eq: quasi stationary distribution}
   f(\omega \, | \, P ) = \frac{1}{Z} \times 
    \begin{cases}
    \displaystyle \exp \left( \frac{2 P }{\epsilon^2} \omega\right) \quad &\omega \in \Omega_0\\[1em]
    \displaystyle C_1 \exp \left( - \frac{\gamma_1}{\epsilon^2} \left( \omega - \Lambda_1 \right)^2  \right) &\omega \in \Omega_1 \\[1em]
    \displaystyle C_2  \exp \left( - \frac{\gamma_2}{\epsilon^2} \left( \omega - \Lambda_2 \right)^2  \right) &\omega \in \Omega_2 \, ,
    \end{cases}
\end{align}
where
\begin{align}
    \label{eq: gaussian shift 1}
    \Lambda_1&= \sign\omega \omega_{0} + {P}/{\gamma_1} \\
    \label{eq: gaussian shift 2}
    \Lambda_2 &= \sign\omega \omega_{1} - ({\gamma_1}/{\gamma_2}) \, \sign\omega (\omega_{1} - \omega_{0}) + {P}/{\gamma_2} \,.
\end{align}
The scaling factors $C_1$ and $C_2$ ensure continuity (see the literature for full expressions \cite{lonardi2026understanding}). These factors, together with $\Lambda_1$ and $\Lambda_2$, are constant for $\omega < 0$, $\omega > 0$, and therefore depend only on $P$ and the sign of $\omega$. 

Knowing \Cref{eq: quasi stationary distribution} without normalization is sufficient to perform the integral in \Cref{eq: frequency distribution} numerically. To do so, we construct a uniform mesh of values $(\omega_m, P_n)$ with step sizes $\Delta \omega$ and $\Delta P$, covering the support of the inferred and measured distributions in \Cref{fig: inference results}(b), (e),  and \Cref{fig: frequency fits}. We choose $\Delta \omega$ and $\Delta P$ such that the power and frequency distributions are discretized into 1000 and 500 uniform bins, respectively. For each fixed $P_n$, we approximate the partition function numerically as
\begin{equation}
    Z(P_n)
    \approx
    \sum_m \exp( - \Phi( \omega_m \mid P_n)) \, \Delta \omega \,,
\end{equation}
which lets us estimate the normalized conditional distribution $f(\omega \mid P)$. The stationary frequency distribution also follows from numerical quadrature over the mesh, namely
\begin{equation}
p(\omega_m)
\approx
\sum_n f(\omega_m \mid P_n) \,\varphi(P_n) \, \Delta P \,.
\end{equation}

For numerical integrals, we use Simpson's rule. Additionally, for numerical stability, all computations are performed in log-space. In particular, we evaluate the partition function using LogSumExp to avoid overflow due to large negative values of the potential $\Phi(\omega \mid P)$.

\section{Implementation details}
\label{sec: implementation details}

The main implementation choices concern the algorithm's termination rule and initialization.

We stop the parameter updates when the relative absolute change between successive entries of $\boldsymbol{\theta}$ falls below a small threshold, $\delta = 10^{-6}$. We also impose an upper limit of $10^4$ steps. Once either criterion is met, the algorithm outputs the current parameter values and the corresponding log-likelihood, which are taken as best estimates.

For the initialization, we note that while the least-squares objective in \Cref{eq: least squares obj} is convex for fixed $\boldsymbol{\theta}$, the full objective in \Cref{eq: mle objective} is non-convex and therefore admits multiple local optima at which the descent routine may terminate. In practice, however, we find that the algorithm is not highly sensitive to initialization for both GB and SA data. We perform a validity check over 24-hour-long batches covering the first week of data, choosing, for each batch, $125$ initializations of $\boldsymbol{\theta}$ over a uniform grid compatible with previous parameter estimates \cite{drewnick2025analyzing}:
\begin{align}
\gamma_1 &\in \{0.05 + 0.025n : n = 0, \dots, 4\} \\\
\gamma_2 &\in \{0.15 + 0.025n : n = 0, \dots, 4\} \\
\epsilon &\in \{0.05 + 0.025n : n = 0, \dots, 4\} \,.
\end{align}
All other parameters are set as in the main experiments of \Cref{sec: results}.
This validation test shows negligible variations in the log-likelihood at termination and convergence to effectively indistinguishable minima.
Consequently, we use the fixed initialization $\boldsymbol{\theta} = (\gamma_1, \gamma_2, \epsilon) = (0.1, 0.2, 0.01)$, which works well for our purpose.

\section{Cross-validation to mitigate overfitting}
\label{sec: mitigating overfitting}

\begin{figure}[t]
    \centering
    \includegraphics[width=0.97\linewidth]{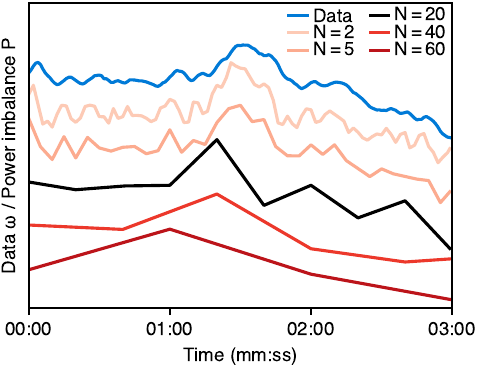}
    \caption{Data and inference results snapshot. We show the first 3 minutes of frequency measurements, along with the inferred power imbalance for different values of $N$. The power imbalance signal shown in black is the one in the main experiments of \Cref{sec: results}. The vertical axis is conventionally scaled. All signals are normalized to unit variance and vertically shifted for visual clarity.}
    \label{fig: overfit snapshot}
\end{figure}

To prevent the inferred power imbalance from fitting measurement noise, we cross-validate our inference algorithm on experimental data. The parameter to tune is the downsampling factor $N$, which controls the temporal resolution of the coarse-grained power imbalance $\tilde{\mathbf P}$ inferred via \Cref{eq: normal equation}. Choosing $N$ too small allows $\tilde{\mathbf P}$ to absorb fast frequency fluctuations that we wish to attribute to noise, whereas choosing $N$ too large prevents $\tilde{\mathbf P}$ from accurately capturing slow variations in the true latent power imbalance. For a visualization of this effect, \Cref{fig: overfit snapshot} shows a representative segment of GB frequency measurements together with the power imbalance inferred at different values of $N$.

\begin{figure*}[htpb]
    \centering
    \includegraphics[width=\linewidth]{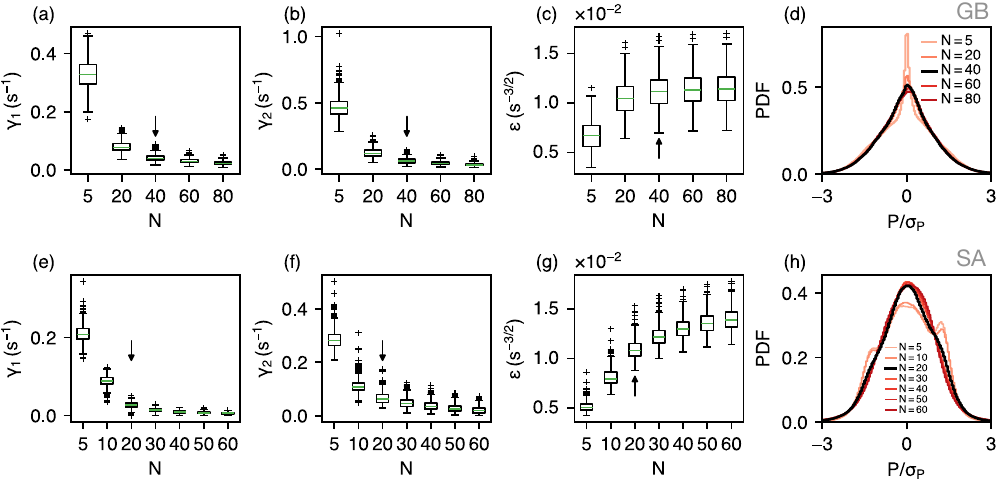}
    \caption{Cross-validation results. Panels (a)-(d) are for the UK, and panels (e)-(h) are for SA. (a)-(c), (e)-(g) Boxplots of the inferred parameters $\boldsymbol{\theta}=(\gamma_1,\gamma_2,\epsilon)$. Green lines indicate the median, while boxes span the interquartile range. Black arrows mark the values of $N$ used in the main experiments; the corresponding parameter values are those reported in \Cref{fig: inference results}(c), (f). (d), (h) Inferred power imbalance distributions, normalized by their standard deviation $\sigma_P$. Black histograms correspond to the values of $N$ used in the main experiments and are shown on a semilogarithmic scale in \Cref{fig: inference results}(b), (e).
}
    \label{fig: overfit}
\end{figure*}

We report cross-validation results in \Cref{fig: overfit}. We test different values of $N$ for GB [\Cref{fig: overfit}(a)-(d)] and SA [\Cref{fig: overfit}(e)-(h)]. Both datasets display qualitatively similar behavior. For small values of $N$, the inferred noise amplitude $\epsilon$ [\Cref{fig: overfit}(c), (g)] is underestimated because the power imbalance absorbs a substantial fraction of the fast fluctuations present in the data. As $N$ increases, the power imbalance parametrization becomes less flexible. Therefore, fast frequency fluctuations can no longer be explained by $\tilde{\mathbf P}$, and the noise amplitude increases toward a plateau. In this latter regime, the noise process becomes more clearly identifiable from the inferred power imbalance.

The damping parameters $\gamma_1$ and $\gamma_2$ [\Cref{fig: overfit}(a)-(b), (e)-(f)] exhibit the opposite trend and decrease with increasing $N$. The reason is that coarser values of $N$ produce a progressively smoother power imbalance, averaging out temporal fluctuations and reducing its amplitude. The inferred dynamics, therefore, require less damping to compensate the power imbalance, leading to smaller values of $\gamma_1$ and $\gamma_2$.

\begin{figure*}[htpb]
    \centering
    \includegraphics[width=1.0\textwidth]{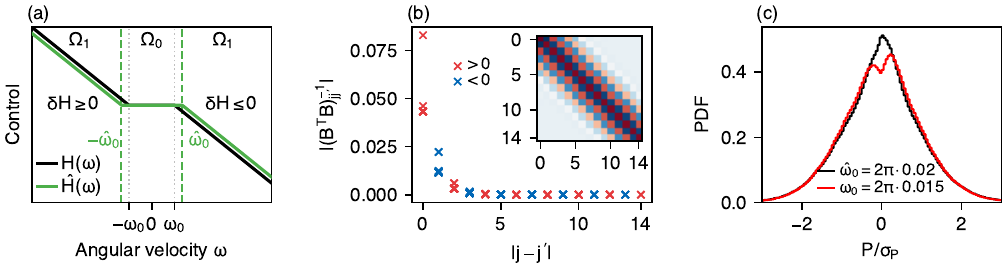}
    \caption{Bias mechanism. (a) Zoom of the control function in \Cref{fig: diagram}(b) around the deadband. Black lines correspond to the nominal control function with deadband size $\omega_0 = 2\pi \cdot \qty{0.015}{rad/s}$, green lines to the control function with enlarged deadband $\hat{\omega}_0 = 2\pi \cdot \qty{0.02}{rad/s}$,  used for inference in the UK. (b) Exponential decay of the entries of the matrix $(\mathbf{B}^{T}\mathbf{B})^{-1}$ as a function of their Manhattan distance from the diagonal (the plot shows a $15 \times 15$ submatrix). Entries in red are positive, while entries in blue are negative. The inset shows the matrix with entries following the same color-sign convention and linearly faded as their magnitudes decay exponentially. (c) Inferred power imbalance distribution with $\omega_0$ in red and $\hat{\omega}_0$ in black (this latter distribution is that shown in \Cref{fig: inference results}(b)).}
    \label{fig: bias}
\end{figure*}

These observations reveal a tension inherent to the model. We therefore select values of $N$ that balance it, namely, values in a regime where the inferred noise amplitude has sufficiently stabilized while the power imbalance retains enough temporal resolution.

Overfitting also manifests itself in the inferred power imbalance distributions [\Cref{fig: overfit}(d), (h)]. For small values of $N$, the inferred power imbalance is contaminated by frequency data and becomes multimodal in SA [\Cref{fig: overfit}(h)]. In GB [\Cref{fig: overfit}(d)], overfitting manifests as a pronounced peak at the center of the distribution. Although one might have expected a multimodal distribution here as well, according to \Cref{fig: frequency fits}(a), a central peak is consistent with the additional bias introduced in the GB inference procedure to further suppress overfitting (see \Cref{sec: biasing the inferred power signal}). As $N$ increases, overfitting-induced features progressively disappear.

\section{Bias in the power imbalance}
\label{sec: biasing the inferred power signal}

For GB data, we slightly enlarge the deadband $\Omega_0$ in \Cref{eq: control 2} by using $\hat{\omega}_0 = 2\pi \cdot \qty{0.02}{rad/s}$, instead of the nominal value $\omega_0 = 2\pi \cdot \qty{0.015}{rad/s}$, during inference. We then restore the deadband to its nominal size when fitting the frequency distribution, namely, for the integral in \Cref{eq: frequency distribution}.

We find that this mild model misspecification mitigates data leakage (overfitting) that cannot be removed by tuning $N$ alone. In \Cref{sec: mitigating overfitting}, we show that choosing an appropriate value of $N$ is key to preventing fitting noise. However, in the UK, we find that setting $N = 20$ together with nominal control boundaries is still insufficient. Indeed, the inferred power imbalance distribution inherits a suppression in the deadband from the frequency distribution [\Cref{fig: bias}(c)]. That is, inference suffers from data leakage.

To mitigate this issue, increasing $N$ is not a viable option, as it leads to an excessive coarsening of the power imbalance and, in turn, to underestimating the damping coefficients [\Cref{fig: overfit}(a), (b)]. We therefore resort to a different solution.

Since during regular grid operation, power imbalance and control approximately compensate each other, we expect the central region of $\varphi(P)$, where $P \approx 0$, to be populated by values inferred when control is idle, i.e., frequency is within the deadband. The comparatively limited number of measurements in the deadband in GB [\Cref{fig: frequency fits}(a), central suppression], however, biases inference, leading to a dip in near-zero values in the reconstructed distribution $\varphi(P)$. This is undesirable, as we do not wish the power imbalance to inherit control features; here, the central suppression of the distribution. By enlarging the deadband size, we attribute idle control to a subset of near-$\omega_0$ yet still-damped fluctuations that formally belong to $\Omega_1$. As a result, we fill the central portion of the inferred power imbalance distribution $\varphi(P)$. We quantify this effect below.

During inference, we fit the frequency data with a new control function $\hat{\mathbf{H}}$ in \Cref{eq: euler maruyama discretization}, instead of $\mathbf{H}$, where $\omega_0$ has been substituted by $\hat{\omega}_0$. Following \Cref{eq: residual}, the residuals that cannot be explained by control then become
\begin{equation}
    \label{eq: residual hat}
    \hat{\mathbf{r}} = \boldsymbol{\Delta \omega} - \Delta t\,\hat{\mathbf{H}} \,.
\end{equation}
Adding and subtracting $\Delta t \, \mathbf{H}$ from $\hat{\mathbf{r}}$ yields
\begin{equation}
    \hat{\mathbf{r}} = \mathbf{r} + \Delta t \, \boldsymbol{\delta}\mathbf{H} \,,
\end{equation}
where $\boldsymbol{\delta} \mathbf{H} = {\mathbf{H}}- \hat{\mathbf{H}}$. 

As per \Cref{eq: normal equation}, the coarse power signal $\hat{\mathbf{P}}$ returned by MLE solves a linear system whose right-hand side is now determined by $\hat{\mathbf{r}}$ instead of ${\mathbf{r}}$. Its solution $\hat{\mathbf{P}}$ reads
\begin{equation}
    \label{eq: bias power}
   \hat{\mathbf{P}} = \tilde{\mathbf{P}} + 
  (\mathbf{B}^T\mathbf{B})^{-1} \mathbf{B}^T \boldsymbol{\delta}\mathbf{H} \,.
\end{equation}
That is to say, if $\boldsymbol{\delta}\mathbf{H} = 0$, namely $\hat{\mathbf{H}} = \mathbf{H}$, then $\hat{\mathbf{P}} = \tilde{\mathbf{P}}$ in the left-hand side of \Cref{eq: bias power}. Conversely, if $\boldsymbol{\delta}\mathbf{H} \neq 0$, then a bias $\mathbf{b} \in \mathbb{R}^M$ is introduced, which is precisely the rightmost term of \Cref{eq: bias power}.

Component-wise the bias depends on $\delta H_k \leq 0$ when $\omega_k \geq 0$ and $\delta H_k \geq 0$ when $\omega_k \leq 0$ [\Cref{fig: bias}(a)], and its entries are
\begin{align}
    \label{eq: bias index}
    b_j
    &=
    \sum_k w_{jk} \,\delta H_k \\
    w_{jk} &= 
    \sum_{j'} (\mathbf{B}^T \mathbf{B})^{-1}_{jj'} \, B_{kj'} \,.
\end{align}
We recall that $j$ and $j'$ are coarse-grid indices and $k$ is a fine-grid index.
By construction (see \Cref{ssec: model discretization}), the $k$-th row of $\mathbf{B}$ has at most two nonzero entries, corresponding to the neighboring coarse-grid columns $j(k)$ and $j(k)+1$, where $j(k)$ is defined by the condition $t_k \in [t_{j(k)}, t_{j(k)+1}]$. The interpolation weight $\lambda_k \in [0,1]$ specifies the relative position of the fine-grid time point $t_k$ within this coarse interval, quantifying how close $t_k$ is to the endpoints $t_{j(k)}$ and $t_{j(k)+1}$. Consequently, the weights $w_{jk}$ take the form
\begin{equation}
    \label{eq: weights explicit}
    w_{jk}
    =
    (1-\lambda_k) (\mathbf{B}^T \mathbf{B})^{-1}_{j,\,j(k)}
    +
    \lambda_k (\mathbf{B}^T \mathbf{B})^{-1}_{j,\,j(k)+1} \,.
\end{equation}

Although $(\mathbf B^T\mathbf B)^{-1}$ is not generically guaranteed to be entrywise positive, its entries are bounded in magnitude. Indeed, since $\mathbf B^T\mathbf B$ is symmetric positive definite and tridiagonal, there exist $C>0$ and $0<\rho<1$ such that $|(\mathbf{B}^T \mathbf{B})^{-1}_{jj'}|
\leq C \rho^{|j-j'|}$ \cite{demko1977inverses}. By direct inspection, we observe that diagonal contributions of $(\mathbf B^T \mathbf B)^{-1}$ are positive, while off-diagonal contributions decay exponentially in magnitude and exhibit alternating signs, as shown in \Cref{fig: bias}(b). Retaining only diagonal contributions $\alpha_j = (\mathbf B^T \mathbf B)^{-1}_{jj}$ in \Cref{eq: weights explicit}, we obtain the rough approximation
\begin{equation}
\label{eq: weights leading order}
    w_{jk}
    \approx
    \begin{cases}
    (1-\lambda_k) \alpha_{j(k)} & \text{if } j = j(k) \\
    \lambda_k \alpha_{j(k)+1} & \text{if } j = j(k)+1 \\
    0 & \text{otherwise} \,.
    \end{cases}
\end{equation}

The outcome of this derivation is: by assuming that positive weights in \Cref{eq: bias index} are well approximated by \Cref{eq: weights leading order}, we expect positive perturbations $\delta H_k$ to produce a negative bias $b_j$, and therefore to shift $\hat{P}_j$ closer to zero compared to $\tilde{P}_j$ in \Cref{eq: bias power}; conversely for negative $\delta H_k$.

Empirically, we observe a behavior consistent with our argument in the GB data where the power imbalance distribution $\varphi(P)$ gets filled in the center, as shown in \Cref{fig: bias}(c).

\section{Details on the distribution fits}
\label{sec: tails of the frequency distributions}

\begin{figure}[b]
    \centering
    \includegraphics[width=0.85\linewidth]{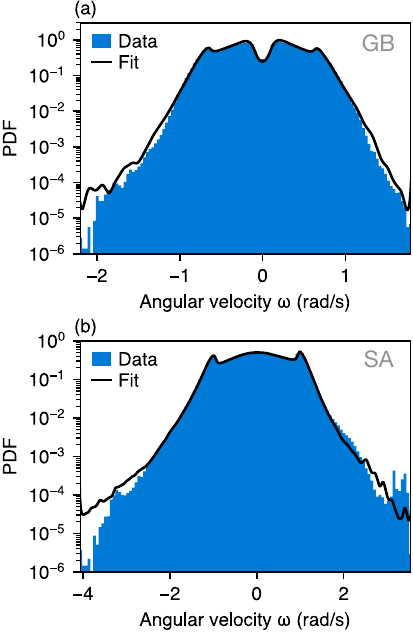}
    \caption{Frequency fit in semilogarithmic scale (companion to \Cref{fig: frequency fits}). Panel (a) is for GB and panel (b) for SA. Black lines are model predictions obtained via \Cref{eq: frequency distribution} while data histograms are in blue.}
    \label{fig: frequency fits log}
\end{figure}

\begin{figure}[t]
    \centering
    \includegraphics[width=0.85\linewidth]{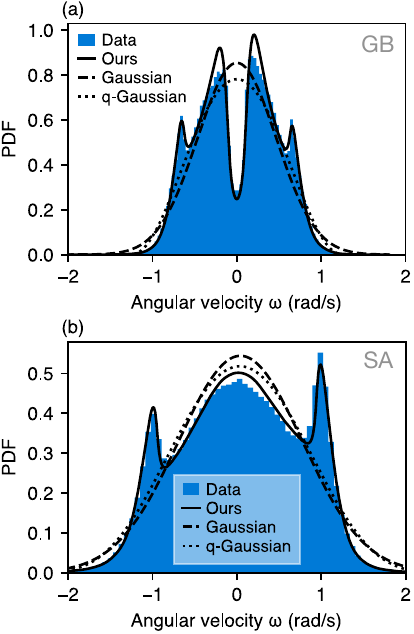}
    \caption{Frequency fit comparison. Panel (a) is for GB and panel (b) for SA. Solid lines are model predictions obtained via \Cref{eq: frequency distribution}, dashed lines report a Gaussian fit, and dotted lines a $q$-Gaussian fit. Data histograms are in blue.}
    \label{fig: frequency fits comparison}
\end{figure}

In \Cref{fig: frequency fits log}, we present a companion to \Cref{fig: frequency fits}, showing the frequency distributions on a semilogarithmic scale. In this way, the heavy tails inherited from $\varphi(P)$ become more apparent. In the tail region, the fits exhibit numerical fluctuations. These arise from the limited number of power imbalance samples available in the tail $\varphi(P)$ for numerical integration (see \Cref{sec: implementation details}). Fluctuations, however, occur only where the frequency probability densities are relatively small and do not affect the overall quality of the fits.

To quantitatively assess the performance of our fitting method, we compare the fit obtained from \Cref{eq: frequency distribution} against Gaussian and $q$-Gaussian distributions, the latter being traditionally used to model the heavy tails of frequency distributions \cite{schaefer2018non}. The comparison is shown in \Cref{fig: frequency fits comparison}. Neither the Gaussian nor the $q$-Gaussian fit fully captures the detailed features of the empirical distributions (GB in \Cref{fig: frequency fits comparison}(a) and SA in \Cref{fig: frequency fits comparison}(b)), suggesting that these models are suited only for coarser representations of the data \cite{schaefer2018non}. In \Cref{tab: nll table}, we also report the negative log-likelihood obtained from the different fitting methods on the empirical data, which indicates stronger statistical support for our model.

For the $q$-Gaussian fits in \Cref{fig: frequency fits comparison}, as well as the tail fits in \Cref{fig: inference results}(b), (e), we use open-source code \cite{githubfoldermle}. In particular, we use the classes \texttt{QGaussian} and \texttt{QGaussianTail}. The former fits a full $q$-Gaussian via MLE, optimizing the parameters $(\mu, q, \beta)$, corresponding respectively to the mean, the non-extensivity parameter, and the shape parameter of the distribution. The latter instead fits only the tails. Here, we fix a cutoff $c > 0$ yielded by the 80th percentile of $|P|$, fold the data by taking absolute values so that both the left and right tails are combined into a single one-sided tail above $c$, and fit the resulting folded tail as a $q$-Gaussian centered at $\mu = c$, optimizing only $(q, \beta)$. The symmetric two-sided tail density is then recovered by mirroring the one-sided fit across zero. This expedient lets us fit the tails with less influence from the bulk of the distribution near $P = 0$. The numerical values of $\beta$ obtained via MLE, complementing those of $q$ reported in the main text, are $\beta = 0.104 \pm 0.001$ in GB and $\beta = 0.251 \pm 0.003$ in SA. Parameters' standard deviations are obtained by inverting the Hessian of the log-likelihood at the optimum \cite{lonardi2026understanding}.

We repeat the fitting procedures with 50 random parameter initializations, selecting the solution with the lowest negative log-likelihood. Finally, for computational efficiency, we perform the fit on a subsample of the measurements $\boldsymbol{\omega}$, obtained via uniform subsampling to achieve a final sample size of $10^6$.

To produce the fits in \Cref{fig: frequency fits} and \Cref{fig: frequency fits log}, we choose the $\boldsymbol{\theta}$ value that best fits the measured data via random-restart hill climbing. We write a detailed pseudocode in \Cref{alg:hillclimb}.

\begin{table}[htpb]
\centering
\caption{Negative log-likelihood values for different fitting methods. Rows are sorted from best (lower negative log-likelihood) to worst. Numbers are rounded to the nearest integer.\vspace{0.5em}}
\label{tab: nll table}
\begin{tabular}{lcc}
\toprule
Method & NLL (UK) & NLL (SA) \\
\midrule
Ours & $\num{12384267}$ & $\num{22490581}$ \\
$q$-Gaussian & $\num{13521083}$ & $\num{23513432}$ \\
Gaussian & $\num{14090777}$ & $\num{23693584}$ \\
\bottomrule
\vspace{1em}
\end{tabular}
\end{table}

\begin{algorithm}[H]
\caption{Random-restart hill climbing}
\label{alg:hillclimb}
\begin{algorithmic}[1]
\State Choose a random parameter $\boldsymbol{\theta} = (\gamma_1,\gamma_2,\epsilon)$ from a batch
\State Initialize $\mathcal{L}_{\mathrm{best}}= + \infty$ and
$n_{\mathrm{stall}}=0$
\For{$\mathrm{step}=1,\ldots,N_{\mathrm{steps}}$} \Comment{$N_{\mathrm{steps}} = 1000$}
    \State Propose a new parameter
    $\boldsymbol{\theta}'$ by randomly shifting entries $\theta_i$ to the previous or next daily batch, or leaving it unchanged
    \State Compute the distribution
    $p(\omega\mid\boldsymbol{\theta}')$ via \Cref{eq: frequency distribution}
    \State Evaluate the nll:
    $
     \mathcal{L}(\boldsymbol{\theta}')
    =
    -
    \sum_{i=1}^{N_{\mathrm{data}}}
    \log p(\omega_i\mid\boldsymbol{\theta}')
    $
    \If{$\mathcal{L}(\boldsymbol{\theta}')
    <
    \mathcal{L}_{\mathrm{best}}$}
        \State Accept the move:
        $\boldsymbol{\theta}\leftarrow\boldsymbol{\theta}'$
        \State Set
        $\mathcal{L}_{\mathrm{best}}
        \leftarrow
        \mathcal{L}(\boldsymbol{\theta}')$, $\boldsymbol{\theta}_{\mathrm{best}}
        \leftarrow
        \boldsymbol{\theta}'$, $n_{\mathrm{stall}} \leftarrow 0$
    \Else
        \State Increment
        $n_{\mathrm{stall}}
        \leftarrow
        n_{\mathrm{stall}}+1$
    \EndIf
    \If{$n_{\mathrm{stall}}
    \geq
    N_{\mathrm{restart}}$} \Comment{$N_{\mathrm{restart}} = 25$}
        \State Randomly resample
        $\boldsymbol{\theta}$ from a batch
        \State Reset
        $n_{\mathrm{stall}} \leftarrow 0$
    \EndIf
\EndFor
\State \Return
$\boldsymbol{\theta}_{\mathrm{best}}$
\end{algorithmic}
\end{algorithm}

\bibliography{bib}

\end{document}